\documentclass[prd,twocolumn,showpacs,nofootinbib,preprintnumbers]{revtex4}
\usepackage{amsmath}
\usepackage{amsfonts}
\usepackage{graphicx}
\usepackage{dcolumn}
\usepackage{hyperref}

\def\be{\begin{equation}}
\def\ee{\end{equation}}
\def\bea{\begin{eqnarray}}
\def\eea{\end{eqnarray}}

\def\5{\overline 5}

\begin{document}
\title{Phantom field dynamics in loop quantum cosmology}

\author{Daris Samart}
\email{jod\_daris@yahoo.com}\vspace{0.6cm}
\affiliation{Fundamental Physics \& Cosmology Research Unit\\ The
Tah Poe Academia Institute (TPTP), Department of Physics\\
Naresuan University, Phitsanulok, 65000 Thailand}

\author{Burin Gumjudpai\footnote{Corresponding author}}
\email{buring@nu.ac.th} \vspace{0.6cm}\affiliation{Fundamental
Physics \& Cosmology Research Unit\\ The Tah Poe Academia
Institute (TPTP), Department of Physics\\ Naresuan University,
Phitsanulok, 65000 Thailand}
\date{\today}

\vskip 1pc
\begin{abstract}
We consider a dynamical system of phantom scalar field under
exponential potential in background of loop quantum cosmology. In
our analysis, there is neither stable node nor repeller unstable
node but only two saddle points, hence no Big Rip singularity.
Physical solutions always possess potential energy greater than
magnitude of the negative kinetic energy. We found that the
universe bounces after accelerating even in the domination of the
phantom field. After bouncing, the universe finally enters
oscillatory regime.
\end{abstract}


\maketitle \vskip 1pc

\section{Introduction} \label{Intro}
Recently, present accelerating expansion of the universe has been
confirmed with observations via cosmic microwave background
anisotropies ~\cite{Spergel:2003cb, Masi:2002hp}, large scale
galaxy surveys~\cite{Scranton:2003in} and type Ia
 supernovae~\cite{Riess:1998cb, Perlmutter:1998np}. However, the problem is that the acceleration can not be understood in
 standard cosmology. This motivates many groups of cosmologists to find out the answers.
Proposals to explain this acceleration made till today could be,
in general, categorized into three ways of approach
\cite{Straumann:2006tv}. In the first approach, in order to
achieve acceleration, we need some form of scalar fluid so called
{\it dark energy} with equation of state $p = w\rho$ where $w <
-1/3$. Various types of model in this category have been proposed
and classified (for a recent review see Ref.
\cite{Copeland:2006wr, Padmanabhan:2004av}). The other two ways
are that accelerating expansion is an effect of backreaction of
cosmological perturbations \cite{Kolb:2005me} or late acceleration
is an effect of modification in action of general relativity. This
modified gravity approach includes braneworld models (for review,
see \cite{Nojiri:2006ri}). Till today there has not yet been true
satisfied explanation of the present acceleration expansion.

Considering dark energy models,  a previous first year WMAP data
analysis combined with 2dF galaxy survey and SN-Ia data and even a
previous SN-Ia analysis alone favor $w < -1$ than cosmological
constant or quintessence \cite{Corasaniti:2004sz, Alam:2003fg}.
Precise observational data analysis (combining CMB, Hubble Space
Telescope, type Ia Supernovae and 2dF datasets) allows equation of
state $p=w\rho$ with constant $w$ value between -1.38 and -0.82 at
the 95 \% of confident level \cite{Melchiorri:2002ux}. The recent
WMAP three year results combined with Supernova Legacy Survey
(SNLS) data when assuming flat universe yields $ -1.06 < w < -0.90
$. However without assumption of flat universe but only combined
WMAP, large scale structure and supernova data implies strong
constraint, $ w = -1.06^{+0.13}_{-0.08}$ \cite{Spergel:2006hy}.
While assuming flat universe, the first result from ESSENCE
Supernova Survey Ia combined with SuperNova Legacy Survey Ia gives
a constraint of $w=-1.07\pm 0.09$ \cite{Wood-Vasey:2007jb}.
Interpretation of various data brings about a possibility that
dark energy could be in a form of phantom field-a fluid with
$w<-1$ (which violates dominant energy condition, $\rho \geq |p|$)
rather than quintessence field
\cite{Caldwell:1999ew,Gibbons:2003yj,Nojiri:2003vn}. The phantom
equation of state $p<-\rho$ can be attained by negative kinetic
energy term of the phantom field. However there are some types of
braneworld model \cite{Sahni:2002dx} as well as Brans-Dicke
scalar-tensor theory \cite{Elizalde:2004mq} and gravitational
theory with higher derivatives of scalar field
\cite{Anisimov:2005ne} that can also yield phantom energy. There
has been investigation on dynamical properties of the phantom
field in the standard FRW background with exponential and
inverse-power law potentials by
\cite{Li:2003ft,Urena-Lopez:2005zd, Hao:2003th, Gumjudpai:2005ry}
and with other forms of potential by \cite{Singh:2003vx,
Sami:2003xv, Gumjudpai:2005ry}. These studies describe fates of
the phantom dominated universe with different steepness of the
potentials.

A problem for phantom field dark energy in standard FRW cosmology
is that it leads to singularity. Fluid with $w$ less than -1 can
end up with future singularity so called the Big Rip
\cite{Caldwell:2003vq} which is of type I singularity according to
classification by \cite{Nojiri:2005sx, Barrow:2004xh}. The Big Rip
singularity corresponds to $a \rightarrow \infty, \rho\rightarrow
\infty $ and $ |p| \rightarrow \infty $ at finite time $t
\rightarrow t_s$ in future. Choosing particular class of potential
for phantom field enables us to avoid future singularity. However,
the avoidance does not cover general classes of potential
\cite{Singh:2003vx}. In addition, alternative model, in which two
scalar fields appear with inverse power-law and exponential
potentials, can as well avoid the Big Rip singularity
\cite{Wei:2005fq}. The higher-order string curvature correction
terms can also show possibility that the Big-Rip singularity can
be absent \cite{Sami:2005zc}.

 Since phantom dominated FRW universe possesses
singularity problem as stated above, in this work, instead of
using standard FRW cosmology, the fundamental background theory in
which we are interested is Loop Quantum Gravity-LQG. This theory
is a non-perturbative type of quantization of gravity and is
background-independent \cite{Thiemann:2002nj, Ashtekar:2003hd}. It
has been applied in cosmological context as seen in various
literatures where it is known as Loop Quantum Cosmology-LQC (for
review, see Ref. \cite{Bojowald:2006da}). Effective loop quantum
modifies standard Friedmann equation by adding a correction term
$-\rho^2/\rho_{\rm lc}$ into the Friedmann equation
\cite{Ashtekar:2006uz, Singh:2006sg, Ashtekar:2006bp, Date:2004zd,
Hossain:2003hb}. When this term becomes dominant, the universe
begins to bounce and then expands backwards. LQG can resolve of
singularity problem in various situations \cite{Ashtekar:2003hd,
Bojowald:2001xe, Ashtekar:2006rx, Ashtekar:2006uz}. However,
derivation of the modified term is under a condition that there is
no matter potential otherwise, in presence of a potential, quantum
correction would be more complicated \cite{Bojowald:2006gr}.
 Nice feature of LQC is avoidance of the future singularity
from the correction quadratic term $-\rho^2/\rho_{\rm lc}$ in the
modified LQC Friedmann equation \cite{Sami:2006wj} as well as the
singularity avoidance at semi-classical regime
\cite{Singh:2003au}. The early-universe inflation has also been
studied in the context of LQC at semi-classical limit
\cite{Bojowald:2003mc, Bojowald:2002nz, Hossain:2003hb,
Tsujikawa:2003vr, Copeland:2005xs, Bojowald:2006hd}. We aim to
investigate dynamics of the phantom field and its late time
behavior in the loop quantum cosmological context, and to check if
the loop quantum effect could remove Big Rip singularity from the
phantom dominated universe. The study could also reveal some other
interesting features of the model.

We organize this article as follows: in section {\ref{SecLQC}}, we
introduce LQC Friedmann equation, after that we briefly present
relevant features of the phantom scalar field in section
{\ref{SecPSF}}. Section {\ref{SecDA}} contains dynamical analysis
of the phantom field in LQC background with exponential potential.
The potential is a simplest case due to constancy of its steepness
variable $\lambda$. Two real fixed points are found in this
section. Stability analysis yields that both fixed points are
saddle points. Numerical results and analysis of solutions can be
seen in section {\ref{numerical}} where we give conditions for
physical solutions. Finally, conclusion is in section
{\ref{SecConc}}.

\section{Loop quantum cosmology} \label{SecLQC}
LQC naturally gives rise to inflationary phase of the early
universe with graceful exit, however the same mechanism leads to a
prediction that present-day acceleration must be very small
\cite{Bojowald:2002nz}. At late time and at large scale, the
semi-classical approximation in LQC formalisms can be validly used
\cite{Bojowald:2001ep}. The effective Friedmann equation can be
obtained by using an effective Hamiltonian with loop quantum
modifications \cite{Singh:2006sg, Sami:2006wj, Singh:2005xg}:
\begin{eqnarray}
\mathcal{C}_{\rm eff}&=&-\frac{3M_{\rm
P}^2}{\gamma^2\bar{\mu}^2}\,a\sin^2(\bar{\mu}\mathfrak{c})+\mathcal{C}_{\rm
m}\,. \label{hamiltonian}
\end{eqnarray}
The effective constraint (\ref{hamiltonian}) is valid for
isotropic model and if there is scalar field, the field must be
free, massless scalar field. The equation (\ref{hamiltonian}),
when including field potential, must have some additional
correction terms \cite{Bojowald:2006gr}. In this scenario, the
Hamilton's equation is
\begin{eqnarray}
 \dot{\mathfrak{p}} &=& \{\mathfrak{p},\mathcal{C}_{\rm eff}\} =
- \frac{\gamma}{3M^2_{\rm P}} \frac{\partial \mathcal{C}_{\rm
eff}}{\partial \mathfrak{c}}\,, \label{Hamilton}
\end{eqnarray}
where $\mathfrak{c}$ and $\mathfrak{p}$ are respectively conjugate
connection and triad satisfying
$\{\mathfrak{c},\mathfrak{p}\}=\gamma/3M_{\rm P}^2$. Dot symbol
denotes time derivative. These are two variables in the simplified
phase space structure under FRW symmetries \cite{Bojowald:2006da}.
Here $M_{\rm P}^2 = (8\pi G)^{-1}$ is square of reduced Planck
mass, $G$ is Newton's gravitational constant and $\gamma$ is
Barbero-Immirzi dimensionless parameter. There are relations
between the two variables to scale factor as $\mathfrak{p}= a^2$
and $\mathfrak{c} = \gamma \dot{a}$. The parameter $\bar{\mu}$ is
inferred as kinematical length of the square loop since its order
of magnitude is similar to that of length. The area of the loop is
given by minimum eigenvalue of LQG area operator.
$\mathcal{C}_{\rm m}$ is the corresponding matter Hamiltonian.
Using the Eq. (\ref{Hamilton}) with constraint from realization
that loop quantum correction of effective Hamiltonian
${\mathcal{C}}_{\rm eff}$ is small at large scale,
$\mathcal{C}_{\rm eff} \approx 0$ \cite{Bojowald:2006da,
Ashtekar:2006bp,Singh:2006sg,Sami:2006wj}, one can obtain
(effective) modified Friedmann equation in flat universe:
\begin{eqnarray}
H^2&=&\frac{\rho_{\rm t}}{3M^2_{\rm P}}\left(1-\frac{\rho_{\rm
t}}{\rho_{\rm lc}}\right)\,, \label{fr}
\end{eqnarray}
where $\rho_{\rm lc} = \sqrt{3}/(16 \pi \gamma^3 G^2 \hbar)$ is
critical loop quantum density, $\hbar$ is Planck constant and
$\rho_{\rm t}$ is total density.

\section{Phantom Scalar Field} \label{SecPSF}

The energy density $\rho$ and the pressure $p$ of the phantom
field contain negative kinetic term. They are given as
\cite{Caldwell:1999ew}
\begin{eqnarray}
\rho&=&-\frac{1}{2}\dot{\phi}^2+V(\phi)\,,\label{rho}\\
p&=&-\frac{1}{2}\dot{\phi}^2-V(\phi)\,. \label{p}
\end{eqnarray}
The conservation law is
\begin{equation}
\dot{\rho}+3H(\rho+p)=0 \,.\label{fluid}
\end{equation}
Using the Eqs. (\ref{rho}), (\ref{p}) and (\ref{fluid}), we obtain
Klein-Gordon equation:
\begin{equation}
\ddot{\phi} + 3H\dot{\phi} - V' = 0 \,, \label{KG}
\end{equation}
where  $V'\equiv {\rm d}V/{\rm d}\phi$ and the negative sign comes
from the negative kinetic terms. The phantom equation of state is
therefore given by
\begin{equation}
w= \frac{p}{\rho} =
\frac{\dot{\phi}^2+2V}{\dot{\phi}^2-2V}\label{EOS}\,.
\end{equation}
From the Eq. (\ref{EOS}), when the field is slowly rolling, as
long as the approximation, $\dot{\phi}^2 \sim 0$ holds, the
approximated value of $w$ is -1.  When the bound, $\dot{\phi}^2 <
2V $ holds, $w$ is always less than -1.

As mentioned before in sections \ref{Intro} and \ref{SecLQC},
there has not yet been a derivation of effective LQC Friedmann
equation in consistence with a presence of potential. Even though,
the Friedmann background is valid only in absence of field
potential, however, investigation of a phantom field evolving
under a potential is a challenged task. Here we also neglect loop
quantum correction effect in the classical expression of Eqs.
(\ref{rho}) and (\ref{p}) (see Refs. \cite{Bojowald:2006gr} and
\cite{Bojowald:2007yy} for discussion).
%
\section{Dynamical analysis} \label{SecDA}
%
Differentiating the Eq. (\ref{fr}) and using the fluid Eq.
(\ref{fluid}), we obtain
\begin{equation}
\dot{H}=-\frac{(\rho+p)}{2M^2_{\rm
P}}\left(1-\frac{2\rho}{\rho_{\rm lc}}\right)\,.\label{hdot}
\end{equation}
The Eqs. (\ref{fr})\,, (\ref{fluid}) and (\ref{hdot}), in
domination of the phantom field, become
\begin{eqnarray}
H^2&=&\frac{1}{3M^2_{\rm P}}\left(-\frac{\dot{\phi}^2}{2}+V\right)\left(1-\frac{\rho}{\rho_{\rm lc}}\right)\,,\label{friedmanphi}\\
\dot{\rho}&=&-3H\rho\left(1+\frac{\dot{\phi}^2+2V}{\dot{\phi}^2-2V}\right)\,,\label{fluidphi}\\
\dot{H}&=&\frac{\dot{\phi}^2}{2M^2_{\rm
P}}\left(1-\frac{2\rho}{\rho_{\rm lc}}\right)\,.\label{hdotphi}
\end{eqnarray}
We define dimensionless variables following the style of
\cite{Copeland:1997et}
\begin{eqnarray}
X \equiv \frac{\dot{\phi}}{\sqrt{6}M_{\rm P} H}\;,
\;\;Y \equiv \frac{\sqrt{V}}{\sqrt{3}M_{\rm P} H}\;,\;\;Z\equiv \frac{\rho}{\rho_{\rm lc}}\,,\\
\lambda\equiv-\frac{M_{\rm P} V'}{V}\;,
\;\;\Gamma\equiv\frac{V\,V''}{\left(V'\right)^2}\;, \;\;\frac{\rm
d~}{{\rm d}N}\equiv \frac{1}{H}\frac{\rm d~}{{\rm d}t}\,,
\end{eqnarray}
where $N\equiv\ln a\,$ is $e$-folding number. Using new variables
in Eqs. (\ref{EOS}) and (\ref{friedmanphi}), the equation of state
is rewritten as\footnote{The relation $\Omega_{\phi} =
{\rho}/{3H^2 M_{\rm P}^2} = -X^2+Y^2 = 1 $ can not be applied here
since it is valid only for standard cosmology with flat geometry.}
\begin{equation}
w  =  \frac{X^2+Y^2}{X^2-Y^2}\,,  \label{EOSnew} \end{equation}
 where $|X|\neq|Y|$ and the Friedmann
constraint is reexpressed as
\begin{equation} (-X^2+Y^2)(1-Z) = 1\label{constraint}\,.
\end{equation}
Clearly, if $|X|\neq|Y|$, following the Eq. (\ref{constraint}),
then $Z \neq 1$. Using the new defined variables above, Eq.
(\ref{hdotphi}) becomes
\begin{equation}
\frac{\dot{H}}{H^2} = 3 X^2 (1-2Z)\,. \label{hdotphinew}
\end{equation}
The acceleration condition,
\begin{equation}
\frac{\ddot{a}}{a}=\dot{H}+H^2>0\label{acceleratecondition},
\end{equation}
in expression of the new variables, is therefore
\begin{equation}
3X^2 (2Z - 1)  <  1\,.
\end{equation}
Divided by the Eq. (\ref{constraint}), the acceleration condition
under the constraint is
\begin{equation}
\frac{3}{1-(Y^2/X^2)} \left(\frac{1-2Z}{1-Z}\right) <
1\label{acceleratecondition2}\,,
\end{equation}
where the conditions $|X|\neq|Y|$ and $ Z \neq 1$ must hold. As we
consider $Z=\rho/\rho_{\rm lc}$ with $\rho = -(\dot{\phi}^2/2) +
V$, we can write
\begin{eqnarray}
 \frac{\rho_{\rm lc} Z}{3M_{\rm P}^2 H^2} &=& -X^2 + Y^2 \,. \label{ZcondXY}
\end{eqnarray}
With the condition $|X|\neq|Y|$, clearly from Eq. (\ref{ZcondXY}),
we have one additional condition, $Z\neq 0$.
\subsection{Autonomous system}
Differential equations in autonomous system are
\begin{eqnarray}
\frac{{\rm d}X}{{\rm d}N}&=&-3X-\sqrt{\frac{3}{2}}\,\lambda
Y^2-3X^3\left(1-2Z\right)\,,   \label{dx}\\
\frac{{\rm d}Y}{{\rm d}N}&=&-\sqrt{\frac{3}{2}}\;\lambda
XY-3X^2Y\left(1-2Z\right)\,,  \label{dy}\\
\frac{{\rm d}Z}{{\rm d}N}&=&-3Z\left(1+ \frac{X^2+Y^2}{X^2-Y^2}\right)\,, \label{dz}\\
\frac{{\rm d}\lambda}{{\rm
d}N}&=&-\sqrt{6}(\Gamma-1)\lambda^2X\label{dlambda}\,.
\end{eqnarray}
Here we will apply exponential potential,
\begin{equation}
V(\phi)=V_0\exp{(-\frac{\lambda}{M_{\rm P}} \phi)}\label{exp
pot}\,,
\end{equation} to this system. The potential is known to yield power-law inflation in standard
cosmology with canonical scalar field. Its slow-roll parameters
are related as $\epsilon = \eta/2 = 1/P$ where $\lambda =
\sqrt{2/P}\;$ and $ P > 1 $ \cite{Lucchin, Liddle}. Although the
potential has been applied to the quintessence scalar field with
tracking behavior in standard cosmology \cite{Barreiro:1999zs},
the quintessence field can not dominate the universe due to
constancy of the ratio between densities of matter and
quintessence field (see discussion in Ref.
\cite{Copeland:2006wr}). In case of phantom field in standard
cosmology under this potential, a stable node is a scalar-field
dominated solution with the equation of state, $w = -1 -
\lambda^2/3$ \cite{Sami:2003xv, Hao:2003th, Kujat:2006vj}. In our
LQC phantom domination context, from Eq. (\ref{dlambda}), we can
see that for the exponential potential, $\Gamma=1$. This yields
trivial value of ${{\rm d}\lambda/{\rm d}N}$ and therefore
$\lambda$ is a non-zero constant otherwise the potential is flat.
\subsection{Fixed points}
Let $f\equiv{{\rm d}X}/{{\rm d}N}, g\equiv{{\rm d}Y}/{{\rm d}N}$
and $h\equiv{{\rm d}Z}/{{\rm d}N}$. We can find fixed points of
the autonomous system under condition:
\begin{eqnarray}
\left(\;f\,,\;g\,,\;h\,\right)\mid_{(X_{\rm c}\;,\;Y_{\rm
c}\;,\;Z_{\rm c})}=0\,.
\end{eqnarray}
The are two real fixed points of this system: \footnote{The other
two imaginary fixed points $(i, 0, 0)$ and $(-i, 0, 0)$ also
exist. However they are not interesting here since we do not
consider model that includes complex scalar field.}
\begin{eqnarray}
&\bullet&{\rm Point~(a)} :
(\frac{-\lambda}{\sqrt{6}},\,~~\sqrt{1+\frac{\lambda^2}{6}}\,,~0\,)\,, \\
&\bullet&{\rm Point~(b)} : (\frac{-\lambda}{\sqrt{6}},
-\sqrt{1+\frac{\lambda^2}{6}}\,,~0\,) \,.
\end{eqnarray}
\begin{table*}[t]
\begin{center}
\begin{tabular}{|c|c|c|c|c|c|c|c|}
\hline
Name&$X$&$Y$&$Z$& ~Existence~ & Stability &$w$ & Acceleration\\
\hline\hline
(a)&$-\frac{\lambda}{\sqrt{6}}$&$\sqrt{1+\frac{\lambda^2}{6}}$&$0$&
All $\lambda$&~
Saddle point for all $\lambda$~&~$-1-\frac{\lambda^2}{3}$&   For all $\lambda$ (i.e. $\lambda^2 > -2 $) \\
\hline (b)&$-\frac{\lambda}{\sqrt{6}}$&
$-\sqrt{1+\frac{\lambda^2}{6}}$&$0$& All $\lambda$ &~ Saddle point for all $\lambda$ ~&~$-1-\frac{\lambda^2}{3}$&  For all $\lambda$ (i.e. $\lambda^2 > -2 $)\\
\hline
\end{tabular}
\caption{{\small Properties of fixed points of phantom field
dynamics in LQC background under the exponential
potential.}}\label{fixedpoint}
\end{center}
\end{table*}
\begin{figure}[t]
\begin{center}
\includegraphics[width=9cm,height=7.7cm,angle=0]{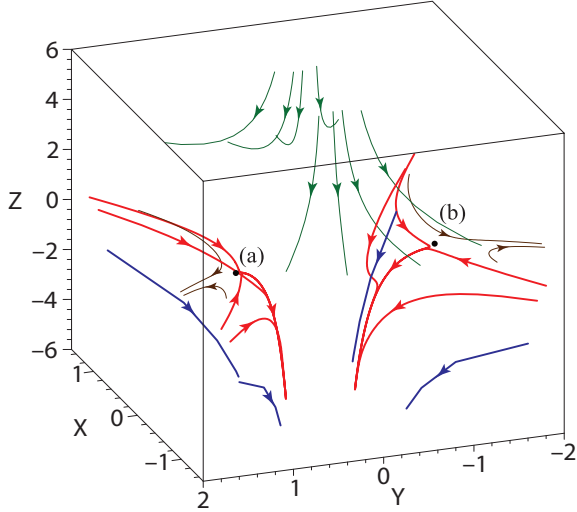}
\end{center}
\caption{Three-dimensional phase space of $X, Y$ and $Z$. The
saddle points (a) (-0.40825, 1.0801, 0)  and (b) (-0.40825,
-1.0801, 0) appear in the figure. $\lambda$ is set to 1. In region
$Z<0$, the solutions (red and blue lines) are non physical. In
this region, $Z\rightarrow-\infty$ when $(X,Y)\rightarrow (0,0)$.
The green lines (class I) are in region $|X|>|Y|$ and $Z > 1$ but
they are also non physical since they correspond to imaginary $H$
values. The only set of physical solutions (class II) is presented
with black lines. They are in region $|Y|>|X|$ and range $0 < Z <
1$. This is the region above (a) and (b) of which $H$ takes real
value. There are separatices $|X|=|Y|$, $Z=0$ and $Z=1$ in the
system (see section {\ref{classII}}).} \label{Phase3D}
\end{figure}
\begin{figure}[t]
\begin{center}
\includegraphics[width=9cm,height=7.2cm,angle=0]{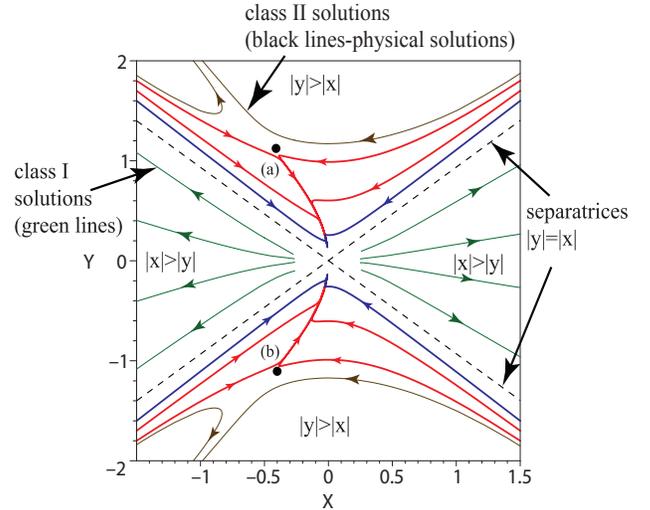}
\end{center}
\caption{Phase space of the kinetic part $X$ and potential part
$Y$ (top view). The saddle points (a) (-0.40825, 1.0801) and (b)
(-0.40825, -1.0801) are shown here. The blue lines and red lines
are in the region $Z<0$ which is non physical. Green lines are of
class I solutions which yields imaginary $H$. Only class II
solutions shown as black lines are physical with real $H$ value.}
\label{Phase2D}
\end{figure}
\begin{figure}[t]
\begin{center}
\includegraphics[width=7.4cm,height=7.3cm,angle=0]{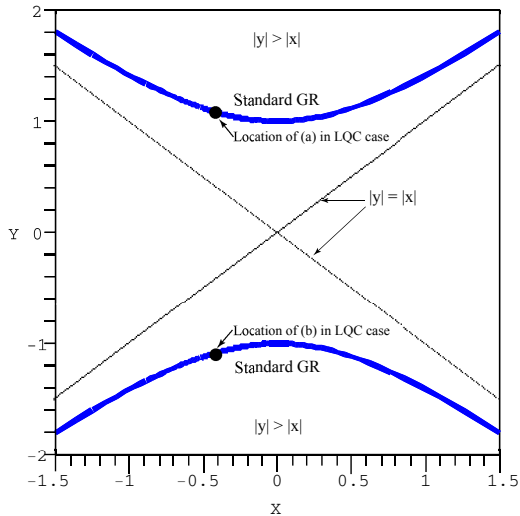}
\end{center}
\caption{Phase space of the kinetic part $X$ and potential part
$Y$ in standard general relativistic case. The location of points
(a) and (b) in Fig. \ref{Phase2D} are on the trajectory solutions
here. This plot shows dynamics of phantom field in standard
cosmological background without any other fluids. In presence of a
barotropic fluid with any equation of state, the point (a) and (b)
correspond to the Big Rip \cite{Urena-Lopez:2005zd,
Gumjudpai:2005ry}. } \label{GRPhase2D}
\end{figure}
\subsection{Stability Analysis} {\label{SecSS}}
Suppose that there is a small perturbation~$\delta X$,~$\delta Y$
and~$\delta Z$ about the fixed point $(X_{{\rm c}}, Y_{{\rm c}},
Z_{\rm c})$, i.e.,
\begin{eqnarray}
X=X_{{\rm c}}+\delta X\,,~~Y=Y_{{\rm c}}+\delta Y\,,~~Z=Z_{\rm
c}+\delta Z\,.
\end{eqnarray}
From Eqs. (\ref{dx}), (\ref{dy}) and (\ref{dz}), neglecting higher
order terms in the perturbations,  we obtain first-order
differential equations:
\begin{eqnarray}
\frac{\rm d }{{\rm d} N} \left(
\begin{array}{c}
\delta X \\
\delta Y \\
\delta Z
\end{array}
\right) = {\cal M} \left(
\begin{array}{c}
\delta X \\
\delta Y \\
\delta Z
\end{array}
\right). \label{uvdif}
\end{eqnarray}
The matrix ${\cal M}$ defined at a fixed point ($X_{\rm c}, Y_{\rm
c},  Z_{\rm c}$) is given by
\begin{eqnarray}
 \label{matM}
{\cal M}=\left( \begin{array}{ccc}
\frac{\partial f}{\partial X}& \frac{\partial f}{\partial Y}& \frac{\partial f}{\partial Z}\\
\frac{\partial g}{\partial X}& \frac{\partial g}{\partial Y}&
\frac{\partial g}{\partial Z}\\
\frac{\partial h}{\partial X}& \frac{\partial h}{\partial Y}&
\frac{\partial h}{\partial Z}
\end{array} \right)_{(X=X_{\rm c}, Y=Y_{\rm c}, Z=Z_{\rm c})}\,.
\end{eqnarray}
We find eigenvalues of the matrix $\mathcal{M}$ for each fixed point: \\\\
$\bullet$ At point (a):
\begin{eqnarray}
\mu_1= \lambda^2 \,,~~\mu_2=-\lambda^2\,,~~\mu_3=
-3-\frac{\lambda^2}{2}\,. \label{eigen c}
\end{eqnarray}
$\bullet$ At point (b):
\begin{eqnarray}
\mu_1= \lambda^2 \,,~~\mu_2=-\lambda^2\,,~~\mu_3=
-3-\frac{\lambda^2}{2} \,.\label{eigen c}
\end{eqnarray}
From the above analysis, each point possesses eigenvalues with
opposite signs, therefore both point (a) and (b) are saddle.
Results from our analysis are concluded in TABLE \ref{fixedpoint}.
Location of the points depends only on $\lambda$ and the points
exist for all values of $\lambda$. Both points correspond to the
equation of state $-1-\lambda^2/3$, that is to say, it has phantom
equation of state for all values of $\lambda \neq 0$. Since there
is no any attractor in the system, a phase trajectory is very
sensitive to initial conditions given to the system. The stable
node (the Big Rip) of the standard general relativistic case in
presence of phantom field and a barotropic fluid, disappears here
(see \cite{Urena-Lopez:2005zd}).

\section{Numerical Results} {\label{numerical}}
Numerical results from the autonomous set (\ref{dx}), (\ref{dy})
and (\ref{dz}) are presented in Figs. \ref{Phase3D} and
\ref{Phase2D} where we set $\lambda=1$. Locations of the two
saddle points are: point (a) ($X_{\rm c} = -0.40825, Y_{\rm c}
=1.0801, Z_{\rm c} =0$) and point (b) ($X_{\rm c} = -0.40825,
Y_{\rm c} =-1.0801, Z_{\rm c} =0$) which match our analytical
results. In Fig. \ref{GRPhase2D}, we present a trajectory solution
of a phantom field evolving in standard cosmological background
for comparing with the trajectories in Fig. \ref{Phase2D} when
including loop quantum effects. The standard case has only simple
two trajectories corresponding to a constraint $-X^2 + Y^2 =1$.
This is attained when taking classical limit, $Z=0$. In loop
quantum case, since there is no any stable node and the solutions
are sensitive to initial conditions, we need to classify solutions
according to each domain region separated by separatrices
$|X|=|Y|$, $Z=0$ and $Z=1$ so that we can analyze them separately.
Note that the condition, $Z > 0$ must hold for physical solutions
since the density can not be negative or zero, i.e. $\rho > 0$.
The blue lines and red lines in Figs. \ref{Phase3D} and
\ref{Phase2D} are solutions in the region $Z<0$ hence are not
physical and will no longer be of our interest. From now on we
consider only the region $Z > 0$. In regions with $|X|>|Y|$, the
solutions therein are green lines (hereafter classified as class
I). The other regions with $|Y|>|X|$ contain solutions seen as
black line (classified as class II). Note that all solutions can
not cross the separatices due to conditions in Eqs.
(\ref{constraint}), (\ref{acceleratecondition2}) and
(\ref{ZcondXY}).)
\subsection{Class I solutions}
Consider the Friedmann equation (\ref{friedmanphi}), the Hubble
parameter, $H$ takes real value only if
\begin{eqnarray}
  \frac{1}{3M^2_{\rm P}}\left(-\frac{\dot{\phi}^2}{2}+V\right)\left(1-\frac{\rho}{\rho_{\rm
  lc}}\right) & \geq & 0\,.
\end{eqnarray}
Divided by $H^2$ on both sides, the expression above becomes
\begin{eqnarray}
  (-X^2 + Y^2)(1-Z) & \geq & 0\,. \label{Hrealcond}
\end{eqnarray}
It is clear from (\ref{Hrealcond}) that, in order to obtain real
value of $H$, class I solutions (green line) must obey both
conditions $|X|>|Y|$ and $Z > 1$ together. However, when imposing
$|X|>|Y|$ to the Eq. (\ref{ZcondXY}), we obtain $Z<0$ instead.
This contradicts to the required range $Z > 1$. Therefore this
class of solutions does not possess any real values of $H$ and
hence not physical solutions.
\subsection{Class II solutions} \label{classII}
\begin{figure}[t]
\begin{center}
\includegraphics[width=8.5cm,height=7.3cm,angle=0]{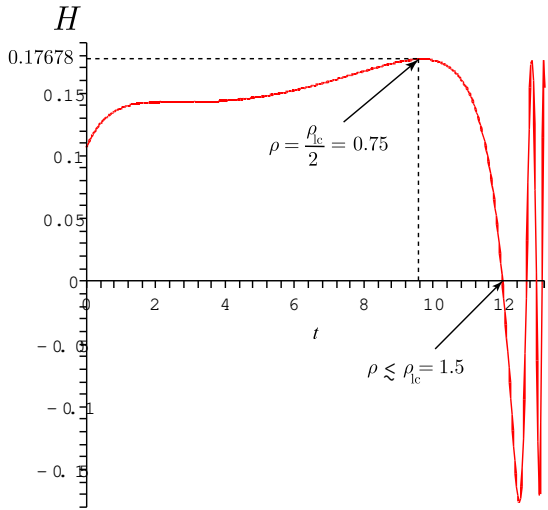}
\end{center}
\caption{Evolution of $H$ with time of a class II solution. Set
values are $\lambda=1,
 \rho_{\rm lc} = 1.5, V_0=1 $ and $M_{\rm P} = 2$. The universe undergoes acceleration from the beginning until
reaching turning point at $\rho=\rho_{\rm lc}/2 = 0.75$ where
$H=H_{\rm max} = 0.17678$. Beyond this point, the universe expands
with deceleration until halting ($H = 0$) at $\rho \approx
\rho_{\rm lc} = 1.5$. After halting, it undergoes contraction
until $H$ bounces. The oscillating in $H$ goes on forever.}
\label{Hvstime}
\end{figure}
\begin{figure}[t]
\begin{center}
\includegraphics[width=8.0cm,height=7.0cm,angle=0]{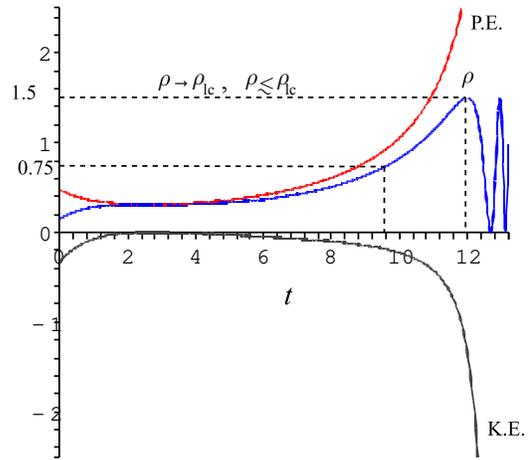}
\end{center}
\caption{Time evolution of potential energy density (P.E.),
kinetic energy density (K.E.)  and $\rho=$ K.E. + P.E. of the
field for a class II solution. K.E. is always negative and, at
late time, it goes to $-\infty$ while P.E. is always positive.
$\rho$ is maximum when $\rho \approx \rho_{\rm lc} = 1.5$. Other
features are discussed as in Fig. \ref{Hvstime}.}
\label{ptdensity}
\end{figure}
\begin{figure}[t]
\begin{center}
\includegraphics[width=8.0cm,height=7.0cm,angle=0]{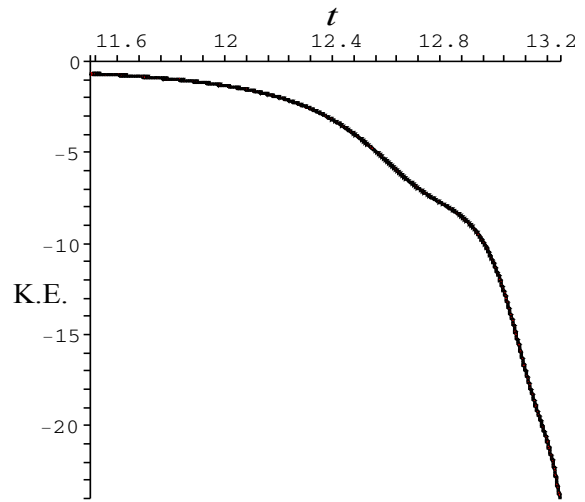}
\end{center}
\caption{Oscillation in kinetic energy density (K.E.) that
contributes to oscillation in $\rho$. This is a zoom-in portion of
the Fig. \ref{ptdensity}.} \label{Kplot}
\end{figure}
Proceeding the same analysis done for class I, we found that in
order for $H$ to be real, class II solutions must obey both
$|Y|>|X|$ and $0 < Z < 1$ together. Moreover when imposing
$|Y|>|X|$ into Eq. (\ref{ZcondXY}), we obtain $Z>0$. Therefore as
we combine both results, it can be concluded that class II
solutions can possess real $H$ value in the region $|Y|>|X|$ and
$0 < Z < 1$, i.e. $0 < \rho < \rho_{\rm lc} $. The bound is
slightly different from the case of canonical scalar field in LQC
(see Ref. \cite{Singh:2006im}) of which the bound is $0 \leq \rho
\leq \rho_{\rm lc} $. The class II is therefore the only class of
physical solutions.

For class II solutions, we consider another set of autonomous
equations from which the evolution of cosmological variables are
conveniently obtained by using numerical approach. In the new
autonomous set, instead of using $N$, which could decrease after
the bounce from LQC effect, time is taken as independent variable.
We define new variable as
\begin{eqnarray}
\dot{\phi}&=& S \,. \label{S}
\end{eqnarray}
The Eqs. (\ref{KG}) and (\ref{hdotphi}) are therefore rewritten as
\begin{eqnarray}
\dot{H} & = & \frac{S^2}{2M_{\rm P}^2} \left[ 1 -
\frac{2}{\rho_{\rm lc}} \left( -\frac{S^2}{2} + V(\phi)
\right)\right]\,, \label{HS} \\ \dot{S} & = & - 3HS+ V'
\,.\label{KGS}
\end{eqnarray}
The Eqs. (\ref{S}), (\ref{HS}) and (\ref{KGS}) form another closed
autonomous system. Numerical integrations from the new system
yield result plotted in Figs. \ref{Hvstime} and \ref{ptdensity} in
which set values are $\lambda=1, \rho_{\rm lc} = 1.5, V_0=1 $ and
$M_{\rm P} = 2$. From Eq. (\ref{fr}) the slope of $H$ with respect
to $\rho$, ${\rm d}H/{\rm d}\rho $, is flat when $ \rho= \rho_{\rm
lc}/2$ \cite{Singh:2006im}. Another fact is
\begin{eqnarray}
\left(\frac{{\rm d}^2{H}}{{\rm d} \rho^2}\right)_{\rho= \rho_{\rm
lc}/2}  =  \frac{-2}{M_{\rm P} \sqrt{3 \rho_{\rm lc}^3}} \;< \;0
\,, \label{Ho2}
\end{eqnarray}
hence, as $ \rho= \rho_{\rm lc}/2$, $H$ takes maximum value,
$H_{\rm max} = \sqrt{\rho_{\rm lc}/12M_{\rm P}^2}$. This result is
valid in LQC scenario regardless of types of fluid.  In Figs.
\ref{Hvstime} and \ref{ptdensity}, with set parameters given
above, as $\rho = \rho_{\rm lc}/2 = 0.75$, $H$ is maximum, $H_{\rm
max} = 0.17678$. When $H \approx 0$, i.e. $\rho$ is approximately
$\rho_{\rm lc} = 1.5$, the expansion halts and then bounces. At
this bouncing point, the dynamics enters loop quantum regime which
is a quantum gravity limit. Beyond the bounce, $H$ turns negative,
i.e. contracting of scale factor. The universe undergoes
accelerating contraction until reaching $H_{\rm min}$. After that
it contracts deceleratingly until bouncing at $H \approx 0$. The
universe goes on faster bouncing forward and backward. The faster
bounce in $H$ is an effect from the faster bounce in $\rho$ as
illustrated in Fig. \ref{ptdensity} where the red line represents
potential energy density $V(\phi)$, the black line represents
kinetic energy density $-\dot{\phi}^2/2$ and the blue line is
total energy density $\rho$.  Oscillation in $\rho$ is from
oscillation in the field speed $\dot{\phi}$ and therefore
oscillation in K.E. as shown in Fig. \ref{Kplot}. This hence
contributes to oscillation in $\rho$. The negative magnitude of
kinetic energy density becomes larger and larger as the field
rolling faster and faster up the potential. The exponential
potential energy density therefore becomes larger and larger.
This results in oscillation of $\rho$ and affects in oscillation
of $H$ about the bounce $H=0$. With a different approach, recently
a similar result in $H$ oscillation is also obtained by Naskar and
Ward \cite{Naskar:2007dn}.

\section{Conclusion} \label{SecConc}
A dynamical system of phantom canonical scalar field evolving in
background of loop quantum cosmology is considered and analyzed in
this work. Exponential potential is used in this system. Dynamical
analysis of autonomous system renders two real fixed points
$(-\lambda/\sqrt{6},\,\sqrt{1+\lambda^2/6}\,,~0\,)$ and $
 (-\lambda/\sqrt{6},
-\sqrt{1+\lambda^2/6}\,,~0\,)$, both of which are saddle points
corresponding to equation of state, $w = -1 - \lambda^2/3$. Note
that in case of standard cosmology, the fixed point $(X_{\rm c},
Y_{\rm c})$ = $(-\lambda/\sqrt{6},\,\sqrt{1+\lambda^2/6})$ is the
Big Rip attractor with the same equation of state, $w = -1 -
\lambda^2/3$ \cite{Hao:2003th}. Due to absence of stable node, the
late time behavior depends on initial conditions given. Therefore
we do numerical plots to investigate solutions of the system and
then classify the solutions. Separatrix conditions $|X|\neq|Y|$,
$Z \neq 1$ and $Z \neq 0$ arise from equation of state
(\ref{EOSnew}), Friedmann constraint (\ref{constraint}) and
definition of $Z$ in Eq. (\ref{ZcondXY}). At first, we consider
solutions in region $Z > 0$, i.e. $\rho > 0$ for physical
solutions. Secondly, within this $Z > 0$ region, we classify
solutions into class I \& II. Solutions in region $|X|>|Y|$ and $Z
>1$ are of class I. However, in order to obtain real value of
$H$ in class I, $Z$ must be negative which contradicts to $Z > 1$.
Therefore the class I solutions are non physical. Class II set is
identified by $|Y|>|X|$ and $ 0 < Z < 1$. It is an only set of
physical solutions since it yields real value of $H$. In class II
set, the universe undergoes accelerating expansion from the
beginning until $\rho=\rho_{\rm lc}/2$ where $H=H_{\rm
max}=\sqrt{\rho_{\rm lc}/12M_{\rm P}^2}$. After that the universe
expands deceleratingly until it bounces, i.e. stops expansion $H
\approx 0$ at $\rho \approx \rho_{\rm lc}$. At the bounce the
universe enters quantum gravity regime. Contraction with backward
acceleration happens right after the bounce, however the
contraction does not go on forever. When the universe reaches
minimum value of negative $H$, the contraction turns decelerated,
i.e. contracts slower and slower down. The universe, after
undergoing contraction to minimum spatial size, bounces again and
starts to expand acceleratingly. Our numerical results yield that
oscillation in $H$ becomes faster as time passes.

\vspace{0.5cm} {\bf Acknowledgements:} We thank Anne-Christine
Davis, Roy Maartens, M. Sami, Shinji Tsujikawa for discussion.
B.~G. thanks Nattapong Yongram for comments to the plot in Fig
\ref{Kplot} and Chakkrit Kaeonikhom for editing some figures.
D.~S. thanks his mother for encouragement. D.~S. is supported by
the Studentship of the Promotion of Science and Mathematics
Talented Teachers Programme of the Institute for the Promotion of
Teaching Science and Technology (IPST). B.~G. expresses his
gratitude to the ICTP, Faculty of Science of Naresuan University
and Suthat Yoksan for opportunity of the ICTP Federation Scheme to
the Abdus Salam ICTP Summer School in Cosmology and Astroparticle
Physics 2006, where partial work was completed. B.~G. is supported
by Faculty of Science of Naresuan University and a TRF-CHE
Research Career Development Grant of the Thailand Research Fund.
Finally, B.~G. has a special thank to the referee for fruitful
comments.


\begin{thebibliography}{99}
\bibitem{Spergel:2003cb}
  C.~L.~Bennett {\it et al.},
  Astrophys.\ J.\ Suppl.\  {\bf 148}, 1 (2003)
  [arXiv:astro-ph/0302207];
D.~N.~Spergel {\it et al.}  [WMAP Collaboration],
Astrophys.\ J.\ Suppl.\  {\bf 148} (2003) 175
[arXiv:astro-ph/0302209].

\bibitem{Masi:2002hp}
  S.~Masi {\it et al.},
  Prog.\ Part.\ Nucl.\ Phys.\  {\bf 48}, 243 (2002)
  [arXiv:astro-ph/0201137].




\bibitem{Scranton:2003in}
R.~Scranton {\it et al.}  [SDSS Collaboration],
[arXiv:astro-ph/0307335].

\bibitem{Riess:1998cb}
  A.~G.~Riess {\it et al.}  [Supernova Search Team Collaboration],
  Astron.\ J.\  {\bf 116}, 1009 (1998)
  [arXiv:astro-ph/9805201];
A.~G.~Riess,
arXiv:astro-ph/9908237;
  J.~L.~Tonry {\it et al.}  [Supernova Search Team Collaboration],
  Astrophys.\ J.\  {\bf 594}, 1 (2003)
  [arXiv:astro-ph/0305008].


\bibitem{Perlmutter:1998np}
  S.~Perlmutter {\it et al.}  [Supernova Cosmology Project Collaboration],
  Astrophys.\ J.\  {\bf 517}, 565 (1999)
  [arXiv:astro-ph/9812133];
G.~Goldhaber {\it et al.}  [The Supernova Cosmology Project
Collaboration],
arXiv:astro-ph/0104382.


\bibitem{Straumann:2006tv}
  N.~Straumann,
  Mod.\ Phys.\ Lett.\ A {\bf 21}, 1083 (2006)
  [arXiv:hep-ph/0604231].

\bibitem{Copeland:2006wr}
  E.~J.~Copeland, M.~Sami and S.~Tsujikawa,
  Int.\ J.\ Mod.\ Phys.\  D {\bf 15}, 1753 (2006)
  [arXiv:hep-th/0603057].


\bibitem{Padmanabhan:2004av}
  T.~Padmanabhan,
  Curr.\ Sci.\  {\bf 88}, 1057 (2005)
  [arXiv:astro-ph/0411044];
  T.~Padmanabhan,
  AIP Conf.\ Proc.\  {\bf 861}, 179 (2006)
  [arXiv:astro-ph/0603114].



\bibitem{Kolb:2005me}
  E.~W.~Kolb, S.~Matarrese, A.~Notari and A.~Riotto,
  arXiv:hep-th/0503117;
  E.~W.~Kolb, S.~Matarrese and A.~Riotto,
  New J.\ Phys.\  {\bf 8}, 322 (2006)
  [arXiv:astro-ph/0506534].


\bibitem{Nojiri:2006ri}
  S.~Nojiri and S.~D.~Odintsov,
  Int.\ J.\ Geom.\ Meth.\ Mod.\ Phys.\  {\bf 4}, 115 (2007)
  [arXiv:hep-th/0601213].


\bibitem{Corasaniti:2004sz}
  P.~S.~Corasaniti, M.~Kunz, D.~Parkinson, E.~J.~Copeland and B.~A.~Bassett,
  Phys.\ Rev.\  D {\bf 70}, 083006 (2004)
  [arXiv:astro-ph/0406608].

\bibitem{Alam:2003fg}
  U.~Alam, V.~Sahni, T.~D.~Saini and A.~A.~Starobinsky,
  Mon.\ Not.\ Roy.\ Astron.\ Soc.\  {\bf 354}, 275 (2004)
  [arXiv:astro-ph/0311364].





\bibitem{Melchiorri:2002ux}
  A.~Melchiorri, L.~Mersini-Houghton, C.~J.~Odman and M.~Trodden,
  Phys.\ Rev.\ D {\bf 68}, 043509 (2003)
  [arXiv:astro-ph/0211522].
\bibitem{Spergel:2006hy}
  D.~N.~Spergel {\it et al.},
  arXiv:astro-ph/0603449.
\bibitem{Wood-Vasey:2007jb}
  W.~M.~Wood-Vasey {\it et al.},
  arXiv:astro-ph/0701041.
\bibitem{Caldwell:1999ew}
  R.~R.~Caldwell,
  Phys.\ Lett.\ B {\bf 545}, 23 (2002)
  [arXiv:astro-ph/9908168].
\bibitem{Gibbons:2003yj}
  G.~W.~Gibbons,
  arXiv:hep-th/0302199.
\bibitem{Nojiri:2003vn}
  S.~Nojiri and S.~D.~Odintsov,
  Phys.\ Lett.\  B {\bf 562} (2003) 147
  [arXiv:hep-th/0303117].
\bibitem{Sahni:2002dx}
  V.~Sahni and Y.~Shtanov,
  JCAP {\bf 0311}, 014 (2003)
  [arXiv:astro-ph/0202346];
  V.~Sahni and Y.~Shtanov,
  Int.\ J.\ Mod.\ Phys.\ D {\bf 11}, 1515 (2000)
  [arXiv:gr-qc/0205111].
\bibitem{Elizalde:2004mq}
  E.~Elizalde, S.~Nojiri and S.~D.~Odintsov,
  Phys.\ Rev.\ D {\bf 70}, 043539 (2004)
  [arXiv:hep-th/0405034].
\bibitem{Anisimov:2005ne}
  A.~Anisimov, E.~Babichev and A.~Vikman,
  JCAP {\bf 0506}, 006 (2005)
  [arXiv:astro-ph/0504560].
\bibitem{Li:2003ft}
  X.~z.~Li and J.~g.~Hao,
  Phys.\ Rev.\ D {\bf 69}, 107303 (2004)
  [arXiv:hep-th/0303093];
  J.~g.~Hao and X.~z.~Li,
  Phys.\ Rev.\ D {\bf 67}, 107303 (2003)
  [arXiv:gr-qc/0302100];

\bibitem{Urena-Lopez:2005zd}
  L.~A.~Urena-Lopez,
  JCAP {\bf 0509}, 013 (2005)
  [arXiv:astro-ph/0507350].
\bibitem{Hao:2003th}
  J.~G.~Hao and X.~z.~Li,
  Phys.\ Rev.\ D {\bf 70}, 043529 (2004)
  [arXiv:astro-ph/0309746].
\bibitem{Gumjudpai:2005ry}
  B.~Gumjudpai, T.~Naskar, M.~Sami and S.~Tsujikawa,
  JCAP {\bf 0506}, 007 (2005)
  [arXiv:hep-th/0502191].
\bibitem{Singh:2003vx}
  P.~Singh, M.~Sami and N.~Dadhich,
  Phys.\ Rev.\ D {\bf 68}, 023522 (2003)
  [arXiv:hep-th/0305110].
\bibitem{Sami:2003xv}
  M.~Sami and A.~Toporensky,
  Mod.\ Phys.\ Lett.\ A {\bf 19}, 1509 (2004)
  [arXiv:gr-qc/0312009].



\bibitem{Caldwell:2003vq}
  R.~R.~Caldwell, M.~Kamionkowski and N.~N.~Weinberg,
  Phys.\ Rev.\ Lett.\  {\bf 91}, 071301 (2003)
  [arXiv:astro-ph/0302506];
  S.~Nesseris and L.~Perivolaropoulos,
  Phys.\ Rev.\ D {\bf 70}, 123529 (2004)
  [arXiv:astro-ph/0410309].

\bibitem{Barrow:2004xh}
  J.~D.~Barrow,
  Class.\ Quant.\ Grav.\  {\bf 21}, L79 (2004)
  [arXiv:gr-qc/0403084].
    \bibitem{Nojiri:2005sx}
  S.~Nojiri, S.~D.~Odintsov and S.~Tsujikawa,
  Phys.\ Rev.\ D {\bf 71}, 063004 (2005)
  [arXiv:hep-th/0501025].

\bibitem{Wei:2005fq}
  H.~Wei and R.~G.~Cai,
  Phys.\ Rev.\ D {\bf 72}, 123507 (2005)
  [arXiv:astro-ph/0509328].


\bibitem{Sami:2005zc}
  S.~Nojiri, S.~D.~Odintsov and M.~Sasaki,
  Phys.\ Rev.\  D {\bf 71}, 123509 (2005)
  [arXiv:hep-th/0504052];
  M.~Sami, A.~Toporensky, P.~V.~Tretjakov and S.~Tsujikawa,
  Phys.\ Lett.\  B {\bf 619}, 193 (2005)
  [arXiv:hep-th/0504154];
  G.~Calcagni, S.~Tsujikawa and M.~Sami,
  Class.\ Quant.\ Grav.\  {\bf 22}, 3977 (2005)
  [arXiv:hep-th/0505193];
  B.~M.~Leith and I.~P.~Neupane,
  arXiv:hep-th/0702002.



\bibitem{Thiemann:2002nj}
  T.~Thiemann,
  Lect.\ Notes Phys.\  {\bf 631}, 41 (2003)
  [arXiv:gr-qc/0210094];
  A.~Perez,
  arXiv:gr-qc/0409061.


\bibitem{Ashtekar:2003hd}
  A.~Ashtekar, M.~Bojowald and J.~Lewandowski,
  Adv.\ Theor.\ Math.\ Phys.\  {\bf 7}, 233 (2003)
  [arXiv:gr-qc/0304074].

\bibitem{Bojowald:2006da}
  M.~Bojowald,
  Living Rev.\ Rel.\  {\bf 8}, 11 (2005)
  [arXiv:gr-qc/0601085].

\bibitem{Date:2004zd}
  G.~Date and G.~M.~Hossain,
  Class.\ Quant.\ Grav.\  {\bf 21}, 4941 (2004)
  [arXiv:gr-qc/0407073].


\bibitem{Ashtekar:2006uz}
  A.~Ashtekar, T.~Pawlowski and P.~Singh,
  Phys.\ Rev.\  D {\bf 73}, 124038 (2006)
  [arXiv:gr-qc/0604013].



\bibitem{Singh:2006sg}
  P.~Singh,
  Phys.\ Rev.\ D {\bf 73}, 063508 (2006)
  [arXiv:gr-qc/0603043].

\bibitem{Ashtekar:2006bp}
  A.~Ashtekar,
  AIP Conf.\ Proc.\  {\bf 861}, 3 (2006)
  [arXiv:gr-qc/0605011].


\bibitem{Hossain:2003hb}
  G.~M.~Hossain,
  Class.\ Quant.\ Grav.\  {\bf 21}, 179 (2004)
  [arXiv:gr-qc/0308014];
  K.~Banerjee and G.~Date,
  Class.\ Quant.\ Grav.\  {\bf 22}, 2017 (2005)
  [arXiv:gr-qc/0501102].

\bibitem{Bojowald:2001xe}
  M.~Bojowald,
  Phys.\ Rev.\ Lett.\  {\bf 86}, 5227 (2001)
  [arXiv:gr-qc/0102069];
  M.~Bojowald, G.~Date and K.~Vandersloot,
  Class.\ Quant.\ Grav.\  {\bf 21}, 1253 (2004)
  [arXiv:gr-qc/0311004];
  G.~Date,
  Phys.\ Rev.\ D {\bf 71}, 127502 (2005)
  [arXiv:gr-qc/0505002].


\bibitem{Ashtekar:2006rx}
  A.~Ashtekar, T.~Pawlowski and P.~Singh,
  Phys.\ Rev.\ Lett.\  {\bf 96}, 141301 (2006)
  [arXiv:gr-qc/0602086]

\bibitem{Bojowald:2006gr}
  M.~Bojowald,
  Phys.\ Rev.\  D {\bf 74}, 081301 (2007)
  [arXiv:gr-qc/0608100];
  M.~Bojowald,
  arXiv:gr-qc/0703144.


\bibitem{Sami:2006wj}
  M.~Sami, P.~Singh and S.~Tsujikawa,
  Phys.\ Rev.\  D {\bf 74}, 043514 (2006)
  [arXiv:gr-qc/0605113].


\bibitem{Singh:2003au}
  P.~Singh and A.~Toporensky,
  Phys.\ Rev.\ D {\bf 69}, 104008 (2004)
  [arXiv:gr-qc/0312110];

\bibitem{Bojowald:2002nz}
  M.~Bojowald,
  Phys.\ Rev.\ Lett.\  {\bf 89}, 261301 (2002)
  [arXiv:gr-qc/0206054].

\bibitem{Bojowald:2003mc}
  M.~Bojowald and K.~Vandersloot,
  Phys.\ Rev.\ D {\bf 67}, 124023 (2003)
  [arXiv:gr-qc/0303072];
  G.~Calcagni and M.~Cortes,
  Class.\ Quant.\ Grav.\  {\bf 24}, 829 (2007)
  [arXiv:gr-qc/0607059].



\bibitem{Tsujikawa:2003vr}
  S.~Tsujikawa, P.~Singh and R.~Maartens,
  Class.\ Quant.\ Grav.\  {\bf 21}, 5767 (2004)
  [arXiv:astro-ph/0311015].
\bibitem{Copeland:2005xs}
  E.~J.~Copeland, J.~E.~Lidsey and S.~Mizuno,
  Phys.\ Rev.\ D {\bf 73}, 043503 (2006)
  [arXiv:gr-qc/0510022].

\bibitem{Bojowald:2006hd}
  M.~Bojowald and M.~Kagan,
  Phys.\ Rev.\  D {\bf 74}, 044033 (2006)
  [arXiv:gr-qc/0606082].





\bibitem{Bojowald:2001ep}
  M.~Bojowald,
  Class.\ Quant.\ Grav.\  {\bf 18}, L109 (2001)
  [arXiv:gr-qc/0105113].

\bibitem{Singh:2005xg}
  K.~Vandersloot,
  Phys.\ Rev.\ D {\bf 71}, 103506 (2005)
  [arXiv:gr-qc/0502082];
 P.~Singh and K.~Vandersloot,
  Phys.\ Rev.\ D {\bf 72}, 084004 (2005)
  [arXiv:gr-qc/0507029].

\bibitem{Bojowald:2007yy}
  M.~Bojowald,
  arXiv:0705.4398 [gr-qc].


\bibitem{Copeland:1997et}
  E.~J.~Copeland, A.~R.~Liddle and D.~Wands,
  Phys.\ Rev.\ D {\bf 57}, 4686 (1998)
  [arXiv:gr-qc/9711068].

\bibitem{Lucchin} F. Lucchin and S. Matarrese, Phys. Rev. D {\bf 32}, 1316 (1985).

\bibitem{Liddle} A. R. Liddle, Phys. Lett. B {\bf 220}, 502 (1989).

\bibitem{Barreiro:1999zs}
  T.~Barreiro, E.~J.~Copeland and N.~J.~Nunes,
  Phys.\ Rev.\ D {\bf 61}, 127301 (2000)
  [arXiv:astro-ph/9910214];
  S.~C.~C.~Ng, N.~J.~Nunes and F.~Rosati,
  Phys.\ Rev.\ D {\bf 64}, 083510 (2001)
  [arXiv:astro-ph/0107321].

\bibitem{Kujat:2006vj}
  J.~Kujat, R.~J.~Scherrer and A.~A.~Sen,
  Phys.\ Rev.\  D {\bf 74}, 083501 (2006)
  [arXiv:astro-ph/0606735].

\bibitem{Singh:2006im}
  P.~Singh, K.~Vandersloot and G.~V.~Vereshchagin,
  Phys.\ Rev.\  D {\bf 74}, 043510 (2006)
  [arXiv:gr-qc/0606032].

\bibitem{Naskar:2007dn}
  T.~Naskar and J.~Ward,
  arXiv:0704.3606 [gr-qc].






\end{thebibliography}
\end{document}